# Profiles of AI Dependency: A Latent Class Analysis of Filipino Students' Academic Competencies

Emerson Quiambao Fernando
College of Education
Pampanga State University
Lubao, Pampanga, Philippines
eqfernando@pampangastateu.edu.ph

Julius Ceazar Gatchalian Tolentino
College of Education
Pampanga State University
Bacolor, Pampanga, Philippines
jcgtolentino@pampangastateu.edu.ph

Maria Anna David Cruz
College of Hospitality and Tourism Management
Pampanga State University
City of San Fernando, Pampanga Philippines
madcruz@pampangastateu.edu.ph

Jordan Lansang Salenga
College of Computing Studies
Pampanga State University
Bacolor, Pampanga, Philippines
jlsalenga@pampangastateu.edu.ph

Vernon Grace Magat Maniago
College of Computing Studies
Pampanga State University
Sto. Tomas, Pampanga, Philippines
vgmmaniago@pampangastateu.edu.ph

Juvy Cruz Grume
College of Computing Studies
Pampanga State University
Bacolor, Pampanga, Philippines
jncruz@pampangastateu.edu.ph

Erika Mamucud Pineda
College of Education
Pampanga State University
Bacolor, Pampanga, Philippines
empineda@pampangastateu.edu.ph

Aileen Pobre De Leon
College of Computing Studies
Pampanga State University
Bacolor, Pampanga, Philippines
apdeleon@pampangastateu.edu.ph

John Paul Palo Miranda
College of Computing Studies
Pampanga State University
Mexico, Pampanga, Philippines
jppmiranda@pampangastateu.edu.ph

## Abstract

The increasing dependency among Filipino college students on artificial intelligence (AI) poses concerns about the potential decline of fundamental academic competencies. This study examines the extent of AI dependency and its perceived effects on students' critical thinking, writing skills, learning independence, research skills, and academic engagement. Using a cross-sectional research design, data was collected from 651 students enrolled in higher education institutions (HEIs) in Pampanga, Philippines accredited by the Commission on Higher Education. The survey data was analyzed using Latent Class Analysis (LCA) to identify AI dependency patterns. Findings indicated that students show moderate to high AI dependency, specifically in research and writing tasks. LCA identified four distinct profiles: highly engaged independent learners, selective AI users, moderate AI users, and AI-dependent learners. Notably, AI-dependent learners demonstrated the weakest academic competencies, with significant dependency on AI-generated outputs. The study highlights the need to foster educational policies that integrate AI literacy while preserving essential academic skills. HEIs must also balance technological advancements with curriculum adaptations to promote critical thinking and ethical use of AI. Future research may explore the longitudinal impacts and intervention strategies to mitigate academic skill erosion caused by AI dependency.

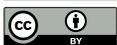

## 1 Introduction

Artificial Intelligence (AI) dependency refers to the increasing reliance on AI tools to perform academic tasks that traditionally require human cognitive effort. In higher education, this dependency has intensified with the rise of generative AI technologies such as ChatGPT. Among Filipino tertiary students, these tools are frequently used for writing, research, and coursework support [22]. While AI provides benefits including real-time feedback, content generation, and grammar enhancement [5], concerns have emerged regarding its potential to erode fundamental academic competencies [3, 6, 10]. Prior research has linked excessive AI use to reduced critical thinking, weakened writing skills, diminished learning independence, and cognitive offloading [8, 12].

Cultural and emotional dimensions further shape AI dependency in the Philippine context. The concept of diskarte, or resourceful problem-solving, may lead students to adopt AI in ways that compromise deeper learning [8]. Additionally, psychological responses such as AI guilt reflect discomfort with issues of authorship, fairness, and academic identity [19]. Although most students recognize the ethical boundaries of AI use and do not equate dependency with dishonesty [2, 3], the growing dependence raises long-term concerns about the decline of essential academic skills and the authenticity of student outputs.

Despite these concerns, existing literature often focuses on general attitudes or isolated academic domains. Few studies have examined AI dependency across multiple academic competencies or used person-centered statistical approaches to identify behavioral profiles in underrepresented regions like the Philippines [23, 24]. This study addresses those gaps by analyzing AI dependency across five domains in Philippine context: critical thinking, writing skills, learning independence, research skills, and academic engagement. Using Latent Class Analysis (LCA), the study identifies distinct profiles of AI use and explores their relationship with academic performance and demographic variables. The study aims to inform educational policy, guide AI literacy efforts, and support balanced integration of AI tools in Philippine higher education.

## 2 Methods
### 2.1 Research Design and Respondents

This study employed a cross-sectional research design to examine the relationship between AI dependency and potential erosion of core academic skills among tertiary students in Pampanga, Philippines. The research was conducted in response to limited published literature on AI dependency and its impact on academic skill development particularly in the context of emerging generative AI tools utilized by students for academic tasks. The study setting, Pampanga, is strategically situated in Central Luzon, Philippines, approximately one hour from Manila, and serves as a major technological and economic hub, hosting the Clark Freeport Zone and Clark International Airport. The province, which comprises 19 municipalities and 2 cities, is home to 45 higher education institutions accredited by the Philippine Commission on Higher Education (CHED). The research gathered data from 800 respondents but only 651 were retained as some of the responses contains incomplete information. These respondents were enrolled in CHED-accredited public and private tertiary institutions across Pampanga which helped the researchers to analyze current student behaviors and perceptions regarding generative AI tool usage in academic settings.

### 2.2 Research Instrument, Data Collection, and Analysis

The study employed a reliable 24-item instrument measuring five dimensions of academic skills: critical thinking, writing skills, learning independence, research skills, and academic engagement. The study was developed by the authors of this study based on existing literature on AI and academic skills. Two educational technology professors reviewed the instrument and provided recommendations. The reliability was assessed using Cronbach's alpha with a separate sample of 30 students who shared characteristics with the target population but were excluded from the main study. The Cronbach's alphas exceeded the 0.70 threshold required for good internal consistency (critical thinking: $\alpha = 0.93$; writing skills: $\alpha = 0.93$; learning independence: $\alpha = 0.97$; research skills: $\alpha = 0.92$; academic engagement: $\alpha = 0.93$) which indicates that the instrument was sufficiently reliable for constructs of interest.

After obtaining institutional approval and informed consent, data was collected using a Google Forms questionnaire from October 2024 to January 2025, which included information on data privacy and security measures in compliance with the Philippine Data Privacy Act. Respondents, recruited through snowball sampling, took 5-15 minutes to complete the questionnaire and were asked to share the link with classmates they knew who used generative AI tools for academic tasks. Data were analyzed using IBM SPSS version 25.0. Descriptive statistics were calculated, and statistical significance was set at $\alpha = 0.05$ for a 95% confidence level. The study utilized LCA to identify distinct student profiles based on their AI dependence and academic skill sacrifice. Previous studies have shown that LCA can be used for such purpose [7, 11]. The LCA model comparison was conducted using Akaike Information Criterion (AIC) and Bayesian Information Criterion (BIC) to determine the optimal number of latent classes that best fit the data [1, 21]. Afterwards, Chi-square analysis was implemented to examine how different demographic factors related to students' AI dependence and academic skill sacrifice patterns. Data was securely stored, processed, and will be disposed of according to protocols.

## 3 Results and Discussion

Descriptive analysis showed that the 651 valid responses came from college students who were 20 years old ($M$ age = 20.22, $SD$ = 1.91), enrolled in both private ($n = 303$) and public ($n = 348$) institutions. The sample predominantly comprised first-year ($n = 248$) and third-year students ($n = 302$), with smaller representations from second-year ($n = 34$) and fourth-year ($n = 67$) levels. Most students self-reported good academic performance ($n = 377$), followed by average ($n = 189$) and excellent ($n = 79$) performance, with very few reporting below average ($n = 5$) or poor ($n = 1$) performance.

Analysis of the five domains revealed concerning patterns in students' AI utilization within academic contexts. The highest dependency scores emerged in three critical areas: uncritical acceptance of AI outputs ($M = 3.23$, $SD = 0.881$), dependency on AI for research suggestions ($M = 3.13$, SD = .909), and substitution of AI for academic engagement ($M = 3.06$, $SD = .914$). While students reported relatively lower dependence on AI for academic decision-making ($M = 2.57$, $SD = .975$) and group participation ($M = 2.60$, $SD = 1.009$), the moderate mean scores across writing skills and learning independence items indicate a troubling shift toward AI-dependent academic practices. The consistency in standard deviations across all 24 items, coupled with higher consensus on specific items, suggests a systematic rather than sporadic pattern of AI dependency. In particular, the overall construct-level means across domains were critical thinking ($M = 2.84$), writing skills ($M = 2.74$), learning independence ($M = 2.72$), research skills ($M$ = 2.90), and academic engagement ($M$ = 2.78). These results indicated a gradual but persistent integration of AI tools as substitutes for fundamental academic processes. The cross-categorical nature of these results demonstrates that new technology dependency like in AI is not confined to isolated academic skills but rather may represent a broad erosion of core academic competencies [17]. Despite maintaining generally satisfactory academic performance, students across various academic years and institutional types are increasingly sacrificing the development of crucial skills in favor of AI-enabled convenience [4]. This trend suggests a potentially detrimental trade-off between short-term efficiency and long-term academic development [4, 24].

Table 1: Student profiles derived from LCA

| Class | Label | CT | WS | LI | RS | AE |
|---|---|---|---|---|---|---|
| 0 | Moderate AI Users | Moderate | Moderate | Somewhat Dependent | Moderate | Moderate |
| 1 | AI-Dependent Learners | Weak | Weak | Very Dependent | Weak | Low |
| 2 | Selective AI Users | Moderate | Strong | Moderate | Strong | High |
| 3 | Highly Engaged Independent Learners | Strongest | Strongest | Strongest | Strongest | High |

Model fit indices indicated that both AIC and BIC values decreased as the number of latent classes increased, with the most substantial improvement observed up to the four-class model (AIC = 109.4, BIC = 4243.13). Beyond four classes, the BIC value began to rise which indicates that risk of overfitting is also rising [1, 16, 21]. Based on these criteria, the four-class solution was identified as optimal which revealed distinct profiles of AI-dependent student behavior (Table 1). Highly Engaged Independent Learners (Class 3) exhibited the strongest academic competencies across all five domains: critical thinking (CT), writing skills (WS), learning independence (LI), research skills (RS), and academic engagement (AE) while showing minimal dependency on AI and using it selectively. Selective AI Users (Class 2) demonstrated strong research and writing skills, moderate independence, and high engagement, adopting AI as a complementary tool rather than a primary resource [9, 13].

Moderate AI Users (Class 0) exhibited a more balanced skill profile, with moderate critical thinking, writing, and research skills, but they showed signs of AI dependence, particularly in learning independence. Although their academic engagement remained moderate, their growing dependency on AI suggested a need for stronger emphasis on verification and analytical reasoning skills. AI-Dependent Learners (Class 1) had the weakest academic competencies across all dimensions, particularly in critical thinking, writing, and research skills. This group demonstrated the highest level of AI dependence, with very low learning independence and engagement. Their substantial dependency on AI for writing, research, and problem-solving indicated a strong tendency to substitute core academic skills with AI-generated content, potentially hindering their ability to develop autonomous learning strategies [18].

Chi-square analyses revealed no significant gender differences in AI dependence patterns ($p$ = .237), suggesting that dependency on AI tools is not influenced by gender. The proportional distribution of male and female students across the latent classes indicates that AI-related academic skill substitution is a broadly shared phenomenon rather than a gender-specific trend [15, 20]. This slightly contrary to earlier studies where men demonstrated a heightened interest in AI chatbots as tools and in their relevance to future career prospects [14]. Similar to gender, academic performance levels did not show a significant association with AI dependence. Both high-achieving and lower-performing students demonstrated comparable levels of AI utilization. This highlights a nuanced relationship: while high-performing students may use AI strategically as a learning enhancement tool, lower-performing students may rely on AI more heavily without experiencing corresponding academic benefits [24]. However, significant variations in AI dependence were observed across academic years ($p$ = .003). Senior students exhibited higher levels of AI dependency, with a greater proportion classified in the high-dependence groups. This trend may be attributed to the increasing academic workload in higher years, coupled with prolonged exposure to advanced AI applications over time [24]. This means that as students' progress academically, their dependency on AI intensifies and may potentially reshape their approach to learning and problem-solving.

## 4 Conclusion, Recommendations, and Implications

This study identified a growing pattern of AI dependency among college students, with AI tools increasingly replacing essential academic skills. LCA revealed four distinct student profiles: Highly Engaged Independent Learners, Selective AI Users, Moderate AI Users, and AI-Dependent Learners. These profiles reflected varying levels of academic competence and dependency on AI. Although AI dependency showed no significant association with gender or academic performance, senior students demonstrated higher levels of dependency, which may be attributed to increased academic demands and extended exposure to AI tools. To address these challenges, institutions should introduce AI literacy programs early in students' academic journeys to encourage responsible use. Educational strategies must focus on strengthening critical thinking, research abilities, and collaborative skills to ensure AI supports rather than substitutes learning. Interventions should be tailored to specific profiles; Moderate AI Users may benefit from training in information validation, AI-Dependent Learners require focused support in skill development, and Highly Engaged Independent Learners can be engaged as peer mentors to promote balanced AI use. Future research should investigate the longitudinal impacts of AI dependency and develop frameworks that integrate AI use while preserving essential academic competencies in a technology-driven educational landscape.